\documentclass{aa} 

\usepackage{graphicx}
\usepackage{txfonts}
\usepackage{orcidlink}
\usepackage{lipsum}
\usepackage{subcaption}         
\usepackage{lscape}             
\usepackage{placeins}           
                                
\hypersetup{
    colorlinks=true,
    linkcolor=blue,
    filecolor=magenta,
    urlcolor=cyan,
    allcolors=blue,
}

\newcommand{\quotes}[1]{``#1''}

\newcommand{\com}{\color{black}}
\def\msun{{\rm M}_\odot}

\begin{document}

   \title{Magnolia: a highly magnetized galaxy at redshift $z\simeq9$}
   \titlerunning{A highly magnetized galaxy at $z\simeq9$}


   \author{E. Ntormousi\inst{1 \orcidlink{0000-0002-4324-0034}}\fnmsep\thanks{Corresponding author: \href{mailto:evangelia.ntormousi@sns.it}{evangelia.ntormousi@sns.it}}
        \and A. Pallottini \inst{2 \orcidlink{0000-0002-7129-5761}}
        \and M. Kohandel\inst{1,3 \orcidlink{0000-0003-1041-7865}}
        \and F. Del Sordo\inst{1,4 \orcidlink{0000-0001-9268-4849}}
        }
   \authorrunning{E. Ntormousi et al.}

   \institute{Scuola Normale Superiore di Pisa, Piazza dei Cavalieri 7, 56126 Pisa, Italy \and Dipartimento di Fisica 'Enrico Fermi', Universit\`{a} di Pisa, Largo Bruno Pontecorvo 3, Pisa I-56127, Italy 
   \and INAF -- Osservatorio di Astrofisica e Scienza dello Spazio di Bologna, via Gobetti 93/3, I-40129 Bologna, Italy \and INAF -- Osservatorio Astrofisico di Catania, via Santa Sofia 78, Catania, Italy}

   \date{Received MMMM DDD, 2026; accepted MMMM DDD, YYYY}

 
  \abstract
   {
   The theory of galactic magnetic field growth via dynamo action predicts that the interstellar medium (ISM) of early galaxies would remain virtually unmagnetized on large scales.
   However, recent observations of strong dust polarization in galaxies at redshifts $z\simeq2.6$ and $z\simeq5.6$ challenge this paradigm. If galaxies indeed host strong, large-scale, coherent magnetic fields at high-$z$, then galaxy evolution and current theories of magnetic field amplification in galaxies should be revisited.
   }
   {
   To address the full physical complexity of early magnetic field growth in galaxies, we have designed MAGNA SERRA, a suite of radiation-magneto-hydrodynamical (RMHD) cosmological zoom-in simulations with non-equilibrium chemistry.
   } 
   {
   In this first paper from the MAGNA SERRA suite, we present Magnolia, a galaxy hosted in a $\simeq 10^{11}~\mathrm{M}_\odot$ dark matter halo simulated with RAMSES-RT with a $\simeq 10^4~\mathrm{M}_\odot$ mass ($\simeq 10 \,\mathrm{pc}$ spatial) resolution. The code is divergence-conserving, giving it a significant advantage over schemes that rely on divergence cleaning.
   }
   {
   We find that, at $z\simeq9$, Magnolia has a stellar mass of a few times $10^8~\mathrm{M}_\odot$ and is strongly magnetized, reaching a magnetic field of a few $\mu G$ 
   and a plasma beta $\beta\simeq 0.1$ in the densest regions.
   This strong amplification occurs over a timescale of about 40~Myrs due to the highly dynamical nature of the galaxy, which creates strong compressions and shear. 
   The galaxy at this redshift is gravitationally unstable, forming clumps that split and merge, creating tidal features in the circumgalactic medium.
   The power spectra of the magnetic and kinetic energies {\com{and the spectral energy transfer functions do not suggest dynamo action, but rather}}
    {\com{an initial}} magnetic field amplification
   {\com{through a large-scale compression, and a later re-organization of this field from small to large scales.}}
   }
   {
   {\com{Our findings show that strong galactic magnetization is possible at very high redshifts, which means that magnetic fields can play a dominant role throughout a galaxy's evolutionary history. Future work, exploring a wide range of possible initial magnetic field conditions, is necessary to make more general statements about early galactic magnetism.}} 
   }


   \maketitle
\nolinenumbers

\section{Introduction}

It is widely accepted that the magnetic field plays a key role in regulating the evolution of present-day, massive, star forming galaxies \citep{Beck1996}.
The magnetic fields typically detected in local spiral galaxies are inferred to be in energy equipartition with the cosmic ray and turbulent energy densities, which means that magnetization must be taken into account when studying the interstellar medium (ISM) dynamics \citep{beck2019}.
Instead, the degree of magnetization of high-redshift galaxies is almost completely unknown, due to the enormous difficulties in measuring magnetic fields even locally \citep{pattle_ppvii}. As a result, the existing estimates for high-redshift magnetization are largely based on theoretical models. 

The most widely accepted theory for sustained magnetic field growth in galaxies, the galactic dynamo, relies on small-scale turbulent motions for amplification and on organized shearing motions, such as differential rotation, for building large-scale coherence \citep{Bran_sub2005,gent2024}.
However, a few Gyrs ($z\lesssim 3-4$) are needed for the large-scale dynamo to amplify the tiny ($\approx \rm nG$) primordial magnetic fields to the typical strengths measured in local galaxies ($\approx \rm \mu G$) and build large-scale coherence.
An alternative could be a rapid amplification of the magnetic field during gravitational collapse \citep[e.g.,][]{Bran_EN2022, Bran_EN2025}, in merger interactions \citep{whittingham2021} or through inflows \citep{Ledos2024}, all of which have only been scarcely explored in terms of their effect on galactic magnetization \citep{Bran_EN23}.

Recently, two measurements of polarized dust emission in the rest-frame FIR of two lensed systems: 9io9, at redshift $z=2.6$ \citep{geach2023}, and SPT0346-52 at redshift $z=5.6$ \citep{chen2024} opened a path in this direction. SPT0346-52 is most probably a merger \citep{litke2019}, while 9io9 is a disk hosting strong turbulence, an ideal environment for dynamo action.
Dust polarization comes from a preferential alignment of elongated dust grains with a strong radiation field or a magnetic field. If the detected dust polarization in these systems indeed comes from the magnetization of their ISM, it means that they host large-scale coherent fields of tens of $\mu$G strength or above. 

This puts a strain on dynamo theories. While 9io9 is at the limit of what dynamo amplification could achieve, the current theoretical paradigm of fast turbulent amplification and subsequent large-scale coherence through a slower, large-scale dynamo does not predict the buildup of a strong, large-scale coherent magnetic field at redshift $z=5.6$.

In this work, we present the first results from MAGNA SERRA, a radiation-magneto-hydrodynamical (RMHD) suite of cosmological zoom-in simulations with non-equilibrium chemistry, designed to study high-redshift galaxies.
MAGNA SERRA includes pre-supernova (SN) feedback in the form of radiation and winds, and also accounts for the photo-chemical evolution of H, He, C, Si, O, and molecular Hydrogen (${\rm H}_2$).
Specifically, here we report the first discovery of a highly magnetized system with an ordered magnetic field, dubbed \quotes{Magnolia}, as early as $z\simeq9$. Our analysis suggests that the significant magnetization of this galaxy happened through a combination of compression and large-scale shear.

\section{Zoom-in cosmological simulations}
The numerical simulations were performed using a customized version of the AMR MHD code RAMSES \citep{teyssier2002,fromang2006} with the extension that includes radiative transfer \citep{Rosdahl2013}. Full details are given in \citet{pallottini:inprep}, while here we report a brief summary.

Our version includes non-equilibrium chemistry through the KROME package \citep{grassi2014}. The network used in this work follows the evolution of H, He, metals (C, Si, O), the formation of ${\rm H}_2$ (a total of 16 species), and their radiative processes \citep{Bovino2016}, which are tracked via 10 distinct radiation bins.
Star formation is based on the ${\rm H}_2$ content of a cell \citep{pallottini2017b}.
{\com{ Specifically, the star formation rate at each cell is calculated using a star-formation efficiency typical of molecular clouds: $\rm \epsilon_\star = 10\%$ \citep[see][]{agertz:2013}, multiplied by the fraction of $\rm H_2$ mass in the cell over the cell's free-fall time. The minimum stellar mass a cell can form is adapted to the mass resolution and, to avoid numerical issues, no more than 90\% of the gas mass can turn into stars in a single event.}}
The stellar chemical, radiative, and energy inputs are pre-computed using the Starburst99 stellar population synthesis \citep{Leitherer1999}. 
Feedback processes like SN explosions and winds are coupled to the gas, which receives an injection of thermal (70\%) and turbulent (30\%) energy: while the former can cool, the latter can be dissipated on an eddy-turnover scale \citep[cf.][]{teyssier:2013, pallottini:2017a}. The turbulent energy, {\com{which we will refer to as "non-thermal" in the following}}, is traced as a separate variable in the simulation. {\com{This variable does not interact with the magnetic field through a sub-grid model, so it would not create an amplification on unresolved scales.}}

The setup for the zoom-in is identical to the simulation in \citet{pallottini2019}. Here we repeat some of the main features for completeness. We use a 20~$\rm Mpc/h$ box from $z\simeq100$. A dark-matter only, coarse cosmological simulation is run to $\rm z=6$ from initial conditions generated by MUSIC \citep{Hahn_Abel2011}.
The target dark matter halo of $10^{11}~M_\odot$ is selected at $\rm z=6$, the simulation is re-run at higher resolution with baryonic physics and, in this case, an initially uniform magnetic field of a $\rm 0.2~nG$ physical strength at redshift z=100 ({\com{$2\times10^{-14}$~G comoving}}).
This initial condition for the magnetic field is compatible with the currently allowed regime for the strength of primordial magnetic seeds on Mpc scales \citep{durrer_neronov, Jedamzik2019, Natwariya2020, Korochkin2021, Pavicevic2025}. 
{\com{Specifically, this value is about 60 times above the lower limit of $\rm B\geq10^{-16}$~G comoving provided by \citep{Neronov2010}, about 20 times above the $\rm B\geq10^{-15}$~G lower limit of \citet{Tavecchio2011}, towards the low end of the allowed regime $\rm 10^{-16}~G\leq B \leq 10^{-9}$~G in \citet{Korochkin2021}, and lower than the Ly$\alpha$ scale-invariant upper limit $B\leq3\times10^{-10}~G$ of \citet{Pavicevic2025}}.}
However, previous work {\com{ \citep[e.g.,][]{martinalvarez2018, katz2021} showed a relative insensitivity of the galactic magnetic fields to the primordial seed.}}

We use the ROCKSTAR-GALAXIES phase-space halo finder \citep{behroozi:2013} to identify halos in each snapshot of the simulation, through which we can reconstruct the entire assembly history of each simulated galaxy \citep[see Sec. 2.4.1 in][]{pallottini:2022}. In this way, we can study their ISM, which in the following will be defined as the properties of the gas inside {\com{a spherical region with radius equal to twice the galaxy's effective radius at a given time\footnote{\com{We have tested that the results remain almost identical if that region is expanded to four times the effective radius, or kept to a fixed value of 2~kpc.}}.}} 

\section{Magnetic field amplification in a high-$z$ galaxy}

\begin{figure*}
    \centering
    \includegraphics[width=0.95\linewidth]{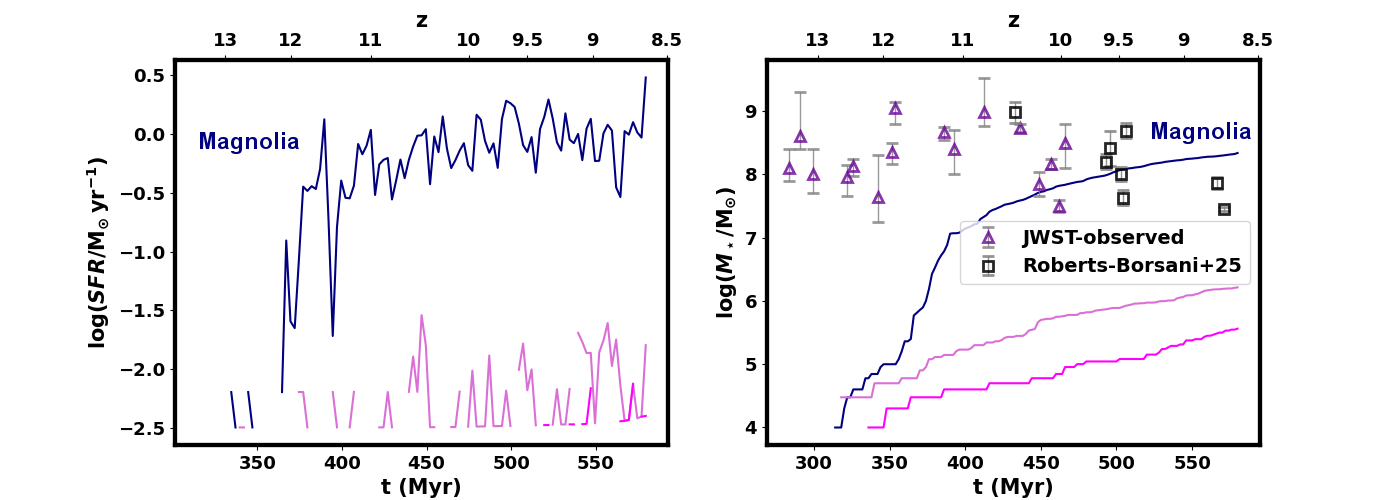}
    \caption{
    Star formation rate (SFR, left) and stellar mass ($M_\star$, right) assembly history of the three galaxies identified in the simulation, as a function of Universe age ($t$, redshift on the upper axis).
    Magnolia is the most massive of the three galaxies, shown in dark blue {\com{and labeled}}. 
    The sample of observations from JWST is reported with black squares \citep{Roberts-Borsani2025} and purple triangles \citep{Carniani2024,robertson2024,fujimoto2024,wang2023,atek2023,curtis_lake2023,castellano2024}.
    \label{fig:mass_assembly_magnolia}    
    }
\end{figure*}

\begin{figure*}
    \centering
    \includegraphics[width=0.49\linewidth]{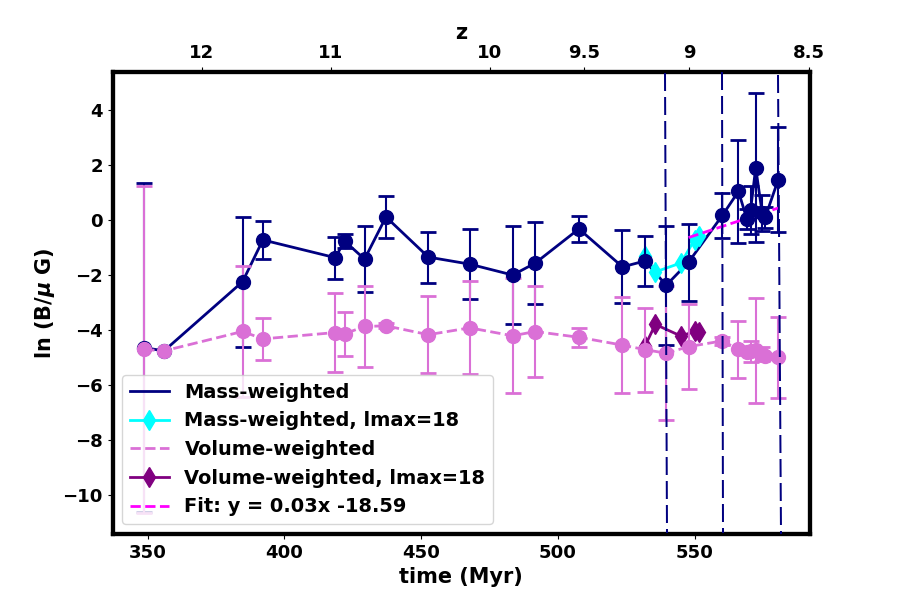}
    \includegraphics[width=0.49\linewidth]{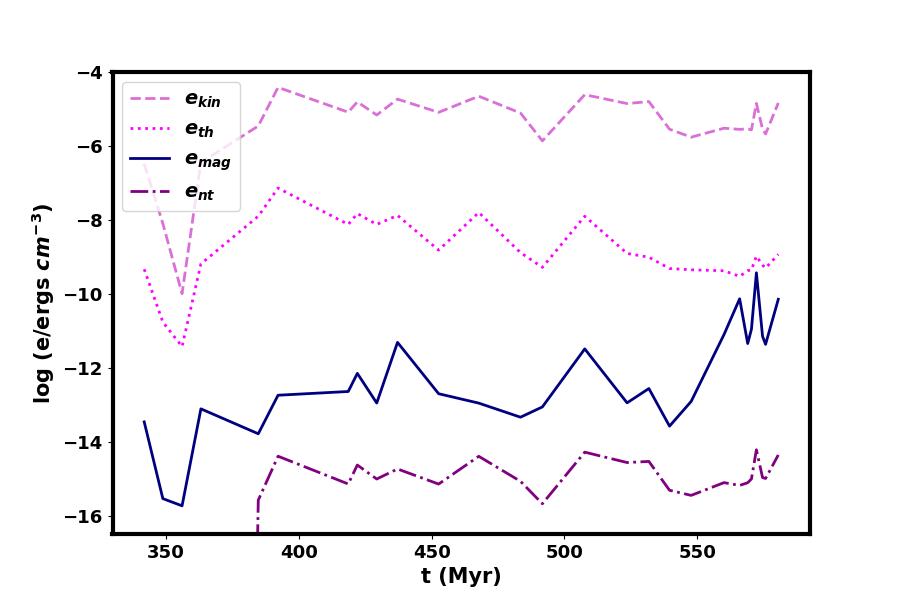}
    \caption{
    \textbf{Left}: Mass-weighted (solid line) and volume-weighted (dashed line) magnetic field as a function of time. The errorbars show the $1\sigma$ variance within the selected volume, and the {\com{dashed blue}} vertical lines denote the three reference snapshots adopted in the subsequent figures and analysis ($z=9.1,8.9,\,{\rm and}\,8.6$).
    {\com{The cyan and purple diamonds show the evolution of the same halo at a lower resolution, and the magenta dashed line is a linear regression fit used to calculate the growth rate over the selected interval.}}
    The shaded pink region shows the magnetic field estimates (upper limit with $3~\sigma$ confidence region) for 9io9 at $\rm z=2.6$ \citep{geach2023} and SPT0346-52 at $\rm z=5.6$ \citep{chen2024}.
    \textbf{Right}: Total mass-weighted energy densities in different components: kinetic ($\rm e_{kin}$), thermal ($\rm e_{th}$), magnetic ($\rm e_{mag}$), {\com{non-thermal}} ($\rm e_{nt}$, see text for details).
    \label{fig:energies_time}
    }
\end{figure*}

\begin{figure*}
    \centering
     \includegraphics[width=0.99\linewidth]{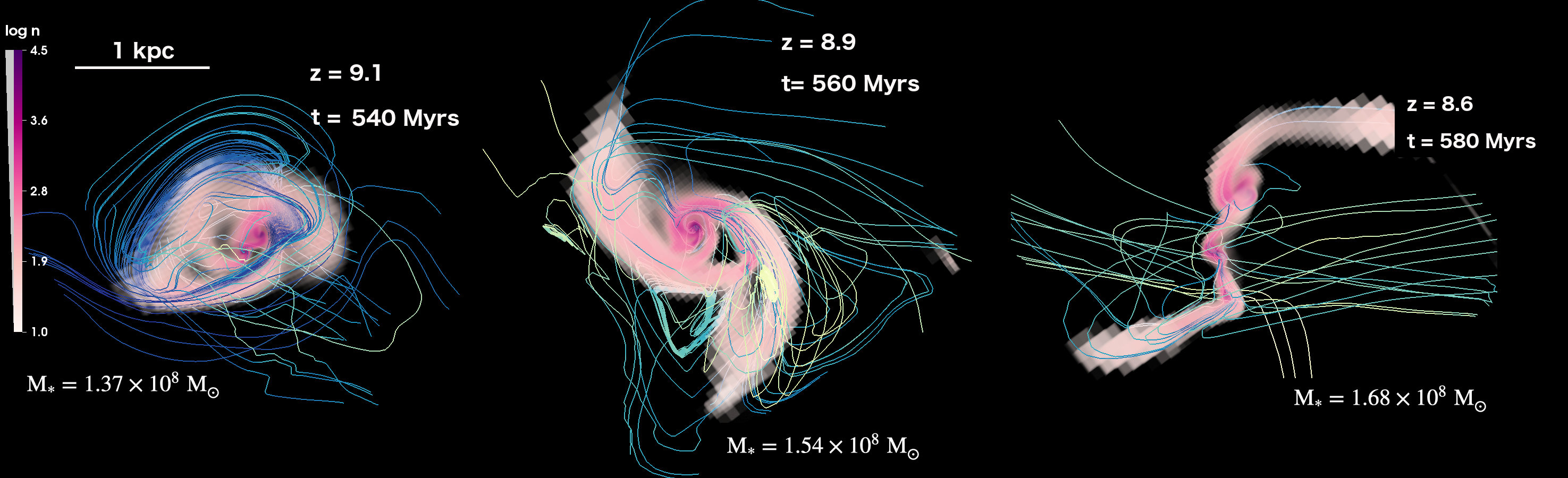}
     \caption{
     Magnolia's density and magnetic field lines during the last 40~Myrs of evolution mark the strong magnetic field amplification.
     The gas number density is shown in dense regions ($\geq 10\, {\rm cm}^{-3}$) as a ray-traced volume rendering based on the column density.
     The magnetic field lines begin at random positions within a 200 pc box centered on the barycenter of the stellar component.
     The field line colors indicate magnetic field strength, from yellow (weakest) to blue (strongest).
     In each figure, we have aligned the view with the principal rotation axis of the stellar component, so that we see the system face-on.
     \label{fig:magnolia_3dlines}
     }
\end{figure*}

\begin{figure*}
    \centering
    \includegraphics[width=0.99\linewidth]{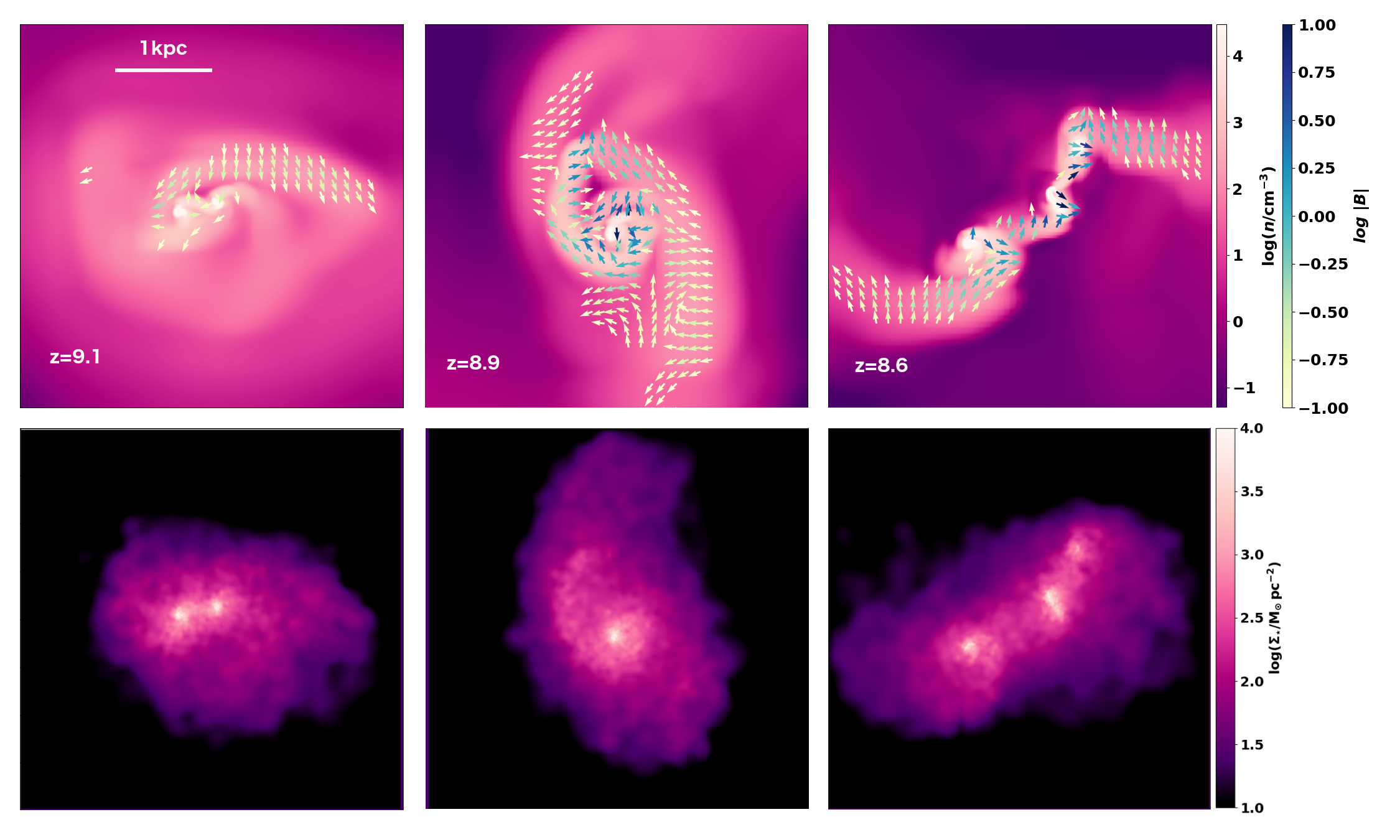}
    \caption{
    Projection of the density and the magnetic field ($n$ and $|\mathbf{B}|$, \textbf{top}) and stellar surface density ($\Sigma_{\star}$, \textbf{bottom}) from $z=9.1$ to $z=8.6$ (left to right).
    The three snapshots are the same as in Fig. \ref{fig:magnolia_3dlines}.
    Projections are along the main stellar rotation axis, and the arrow colors show the projected magnetic field strength.
    \label{fig:projection}
    }
\end{figure*}

The simulation is integrated for about 550~Myrs, up to redshift $z\simeq8.6$. Magnolia is identified as the largest galaxy in the volume, with a stellar mass of about $\rm 8.6\times10^8~M_\odot$ at $z\simeq8.6$.

The star formation rate (SFR) and mass assembly history of Magnolia {\com{and two smaller ($M_\star \lesssim 10^6 \msun$) galaxies identified in the simulation but not analyzed here}} are shown in Fig. \ref{fig:mass_assembly_magnolia}. From the discontinuous rise in stellar mass at $z=12.6$, we notice that the galaxy has undergone a merger around that redshift, while it is growing smoothly until $z=8.6$. Its SFR is of order of 1~$\rm M_\odot~yr^{-1}$ for most of its evolution, and features a moderate burstiness of order of $\sim 0.3$ dex \citep[cfr.][]{pallottini:2023}.
To analyze the interstellar and circumgalactic medium of Magnolia, in the following, we extract all quantities in a sphere of 2~kpc centered on the halo\footnote{A further selection limiting the sampled areas to the densest ($\rm n>0.1~cm^{-3}$) gas only gave almost identical results.}.

The left panel of Fig. \ref{fig:energies_time} shows the evolution of the mass-weighted and volume-weighted magnetic field. 
The mass-weighted magnetic field at the earliest time where the galaxy is identified ($z\simeq13$) is already around $0.1\, \rm \mu G$ and stays almost constant with time, with occasional small (within 1$\sigma$) fluctuations until $\rm z=9$.
Around $z\simeq9.2$ the magnetic field of the galaxy starts increasing, reaching a mean value of almost $\rm 10~\mu G$ at redshift $\rm z=8.6$.
As a reference, pure adiabatic compression ($\rm B\propto\rho^{2/3}$) of the initial ($\rm 2\times10^{-4}~\mu G$) magnetic field from a density of $10^{-4}$ to 1 cm$^{-3}$ would give an average field of $\rm 8\times10^{-2}~\mu G$.
{\com{A log-linear regression fit to the late evolution of the galaxy shows a growth rate of about 30 Gyr$^{-1}$: $\rm |B|\propto e^{0.03 t/Myr}$.}}
This estimate is roughly 15 times higher than the large-scale dynamo growth rate for the Milky Way. In the kinematic regime, and for a flat rotation curve, the rough scaling of the large-scale dynamo growth rate, $\gamma$, is linear with the galaxy's rotation rate: $\gamma\propto\Omega$ \citep[e.g.,][]{Chamandy2014}. So the fact that in Magnolia $\gamma$ is 15 times higher than in the Milky Way implies that, for similar scale height and turbulent diffusivity, Magnolia would have to have a 15 times faster rotational shear, of more than 500~km~s$^{-1}$~kpc$^{-1}$. For this 1~kpc galaxy, this would imply a rotation speed of 1000~km~s$^{-1}$, which is unreasonably high \citep[cfr.][]{kohandel:2024}. Therefore, in this situation, a large-scale dynamo origin of the amplification is implausible.
{\com{As an average value, this growth rate is comparable to that obtained by simulations that resolve a small-scale dynamo \citep[e.g.,][]{pakmor2017,Pfrommer2022,gent2024}. However, as the error bars and abrupt point-to-point variations in Fig. \ref{fig:energies_time} indicate, it fluctuates by roughly an order of magnitude, which is not a characteristic of kinematic dynamos. 
Finally, we overplot a numerical test, i.e. the magnetic field evolution during a restart with one less level of refinement\footnote{Unfortunately, we could only evolve for a limited amount of time due to high CPU cost of the simulation}; the data show a tendency towards a similar growth rate. This fact could indicate that the amplification process is not acting close to the resolution. In the following, we will examine the origin of the amplification more closely.}}

The fact that the volume-averaged field rises smoothly with time while the mass-weighted one is more episodic indicates that the peaks are correlated to dynamical events rather than smooth mass accretion.
Further evidence of that is the continuous mass accretion history of the galaxy, which does not show any major merger after $z\simeq12$ (Fig. \ref{fig:mass_assembly_magnolia}).
Therefore, the peaks in the magnetic field intensity are likely associated with clumps formed in and around the galaxy, occasionally compressing the field. 

To get a better idea of what could be causing the amplification in late times, we turn to the right panel of Fig. \ref{fig:energies_time}, which compares the magnetic energy density to the total kinetic and non-thermal energy injected by supernovae.
While the magnetic energy in Magnolia accounts for only a tiny fraction of the total kinetic energy, it exceeds the non-thermal energy density by two orders of magnitude. It also increases with time up to almost an equipartition value with the thermal energy density. Instead, the non-thermal contribution stays roughly constant with time in line with the SFR, and considering the injection of non-thermal feedback-related energy is balanced by turbulent dissipation.
Instead, the magnetic energy increases.
These elements hint at an amplification driven by compression, large-scale shear, or winding in the forming galaxy rather than supernova-driven turbulence.

\begin{figure*}
    \centering
    \includegraphics[width=0.33\linewidth]{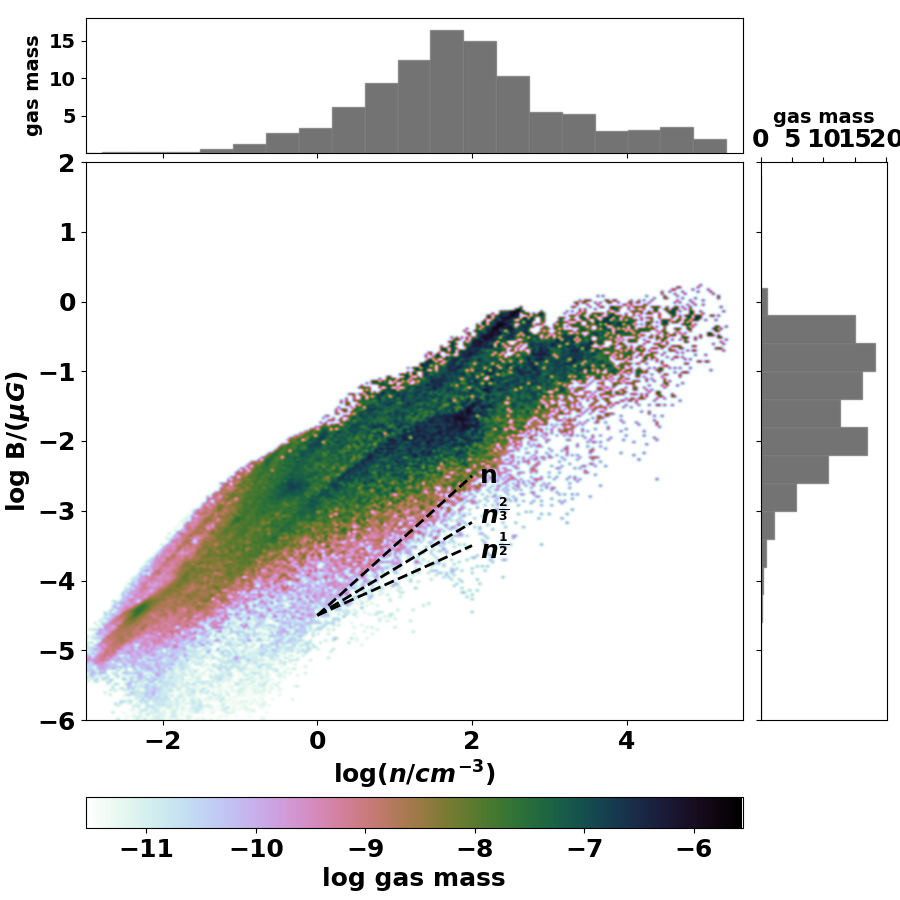}
    \includegraphics[width=0.33\linewidth]{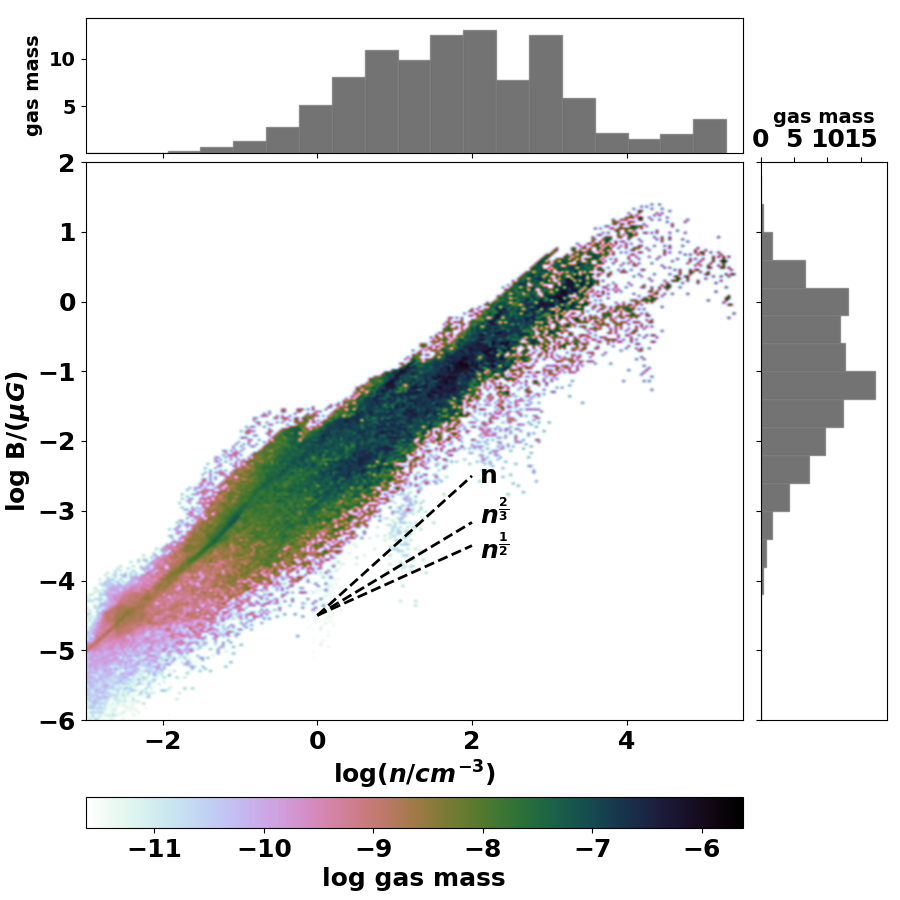}
    \includegraphics[width=0.33\linewidth]{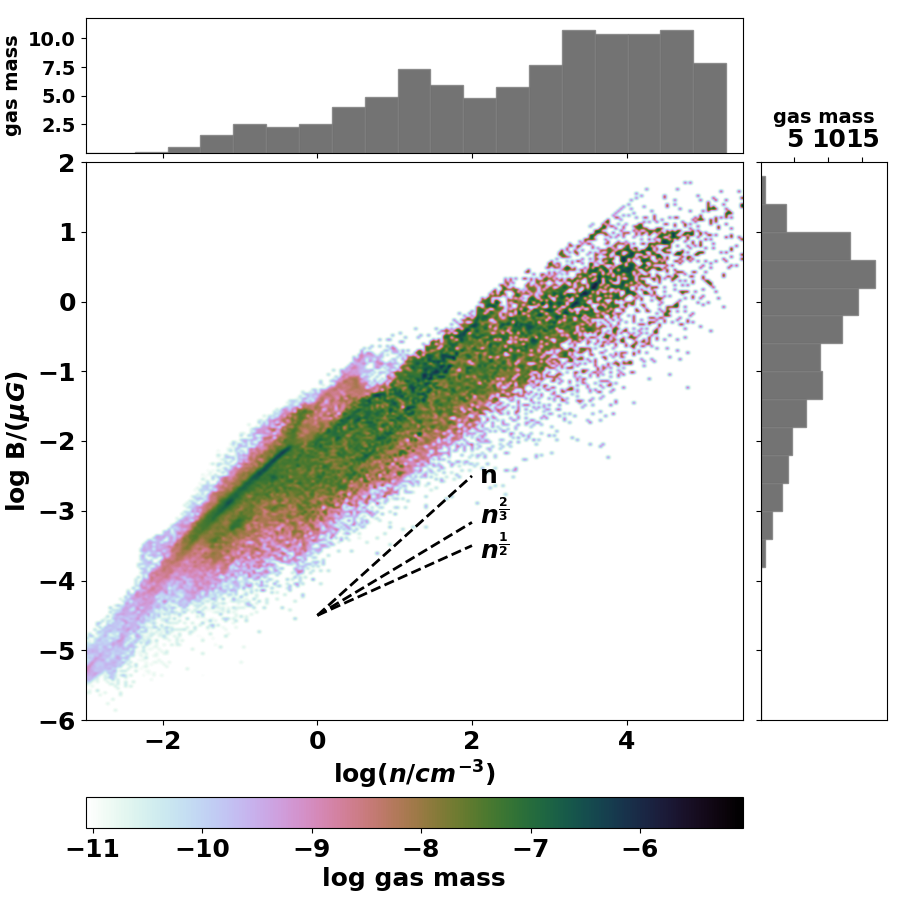}
    \caption{
    Distribution of the ISM in the magnetic field - density plane for Magnolia at $z=9.0$ (right before the strong amplification stage), $z=8.9$ (at the start of the amplification stage), and $z=8.6$ (after the amplification).
    The dashed lines show reference slopes for different magnetic field amplification mechanisms: i) compression along field lines ($\rm B\propto\rho$), spherical collapse ($\rm B\propto\rho^{2/3}$), and compression with constant plasma beta ($\rm B\propto\rho^{1/2}$).
    Note that the distributions are weighted by the gas mass, and we show the 1-D distributions as insets.
    \label{fig:eos_magnetic_field}    
    }

\end{figure*}

The 3D structure of the magnetic field is shown together with the gas density in Fig. \ref{fig:magnolia_3dlines}.
We have chosen three snapshots that roughly correspond to the time before the strong amplification phase ($z=9.1$), its beginning ($z=8.9$), and after the amplification ($z=8.6$). 
The galaxy seems gravitationally unstable, forming clumps that split and merge, also creating prominent, long tidal features.
The magnetic field is very tangled and only follows the density structure in the densest regions. Notably, there are regions of a relatively strong magnetic field in the circumgalactic medium.

The same three snapshots are shown in projection along the main rotation axis in Fig. \ref{fig:projection}.
The top panel shows the averaged gas density with overlaid projected magnetic field vectors. Overall, the projected magnetic field is strongest in the highest density regions.
At the beginning of the amplification phase ($z=8.9$), the magnetic field appears relatively ordered in the outskirts, with hints of an ordered, strong ($\rm \mu G$) field in the outskirts of the galaxy.
After the amplification ($z=8.6$), this effect is more pronounced, with moderate density gas ($n\simeq1\, \rm cm^{-3}$) showing an ordered projected magnetic field of a few $\rm\mu G$ on kpc scales.
From Fig. \ref{fig:magnolia_3dlines} we can confirm that this field is aligned with the elongated density structure around the dense clumps. Instead, in the dense regions ($n\gtrsim 10^2\rm \,cm^{-3}$), the field appears disordered.

The bottom panel of Fig. \ref{fig:projection} shows the corresponding stellar surface density. 
The stars are organized in clumps that merge and split over a few Myr.
{\com{This highly dynamical process is very reminiscent of a merger, albeit on smaller temporal and spatial scales. The frequent tidal interactions between these clumps create strong strains that can have a considerable impact on the magnetic field through the induction equation.}}

Fig. \ref{fig:eos_magnetic_field} shows the magnetic field against the gas density in mass-weighted bins at the same three snapshots as above ($z=9.1$, $8.9$, and $8.6$).
At $z=9$, most of Magnolia's gas hosts magnetic fields between $10^{-3}$ and $1\,\rm \mu G$. Shortly afterwards, at $z=8.9$, the highest-density regions ($n>10^2\, \rm cm^{-3}$) reach magnetic field strengths above $10\, \rm \mu G$. This corresponds to a plasma beta $\beta\simeq 0.1$ in the densest regions, similar to the Milky Way values.
At $z=8.6$, most of the dense gas mass in Magnolia's ISM is strongly magnetized, with magnetic field values above 1 and up to tens of $\rm\mu$G.
The dashed lines in Fig. \ref{fig:eos_magnetic_field} show reference slopes for classic scenarios of magnetic field compression: assuming a roughly uniform magnetic field, $\rm B\propto\rho$ can result from compression across field lines, while $\rm B\propto\rho^{2/3}$ from spherical collapse. The $\rm B\propto\rho^{1/2}$ relation is more general and results from compression under a constant plasma beta \citep{mouschovias_spitzer_1976}.
The gas in Magnolia does not uniquely follow any one of these relations. Instead, there is a large dispersion of about 2 orders of magnitude around one or two main trends that range from superlinear to $\rm B\propto\rho^{1/2}$. Specifically, at the start of the strong amplification phase ($z=8.9$, middle panel), the relation followed by the bulk of the gas is steeper than linear. This {\com{super-linear correlation is not predicted by any scenario of pure collapse. Other processes such as dynamo action, reconnection, or anisotropic compression could produce this behavior, but they cannot be identified by the $\rm B-\rho$ relation alone.}}

\begin{figure}
    \centering
    \includegraphics[width=0.4\textwidth]{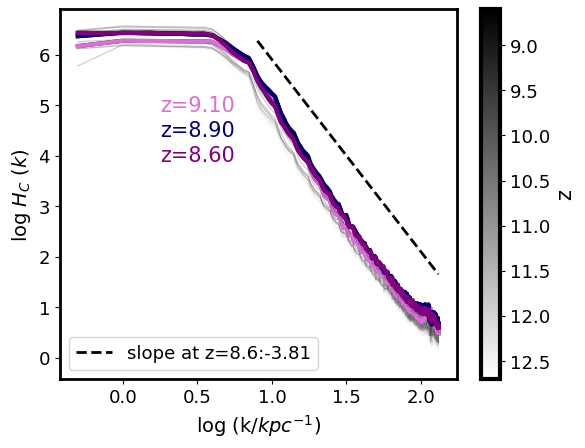}
    \includegraphics[width=0.4\textwidth]{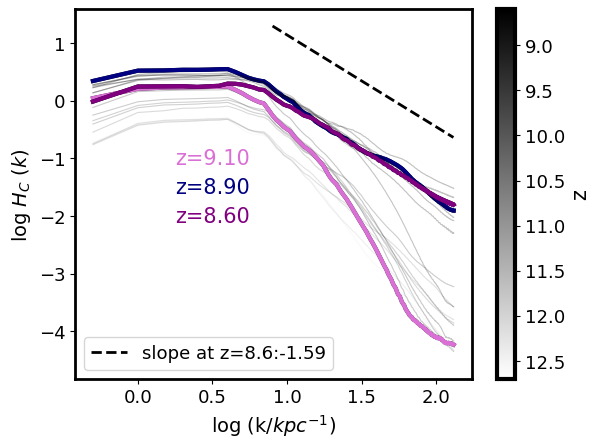}
    \caption{
    Power spectra ($H_C$) of the kinetic (\textbf{top}) and magnetic (\textbf{bottom}) energies as a function of scale for $8.6<z<13$, color-coded by redshift. The wavenumbers are in inverse kpc.
    Snapshots from Fig. \ref{fig:magnolia_3dlines} are shown in contrasting colors, as indicated by the in-plot text. The dashed lines show the power-law fit at $z=8.6$. 
    \label{fig:power_spectra}
    }
\end{figure}

\begin{figure*}
    \centering
    \includegraphics[width=0.8\linewidth]{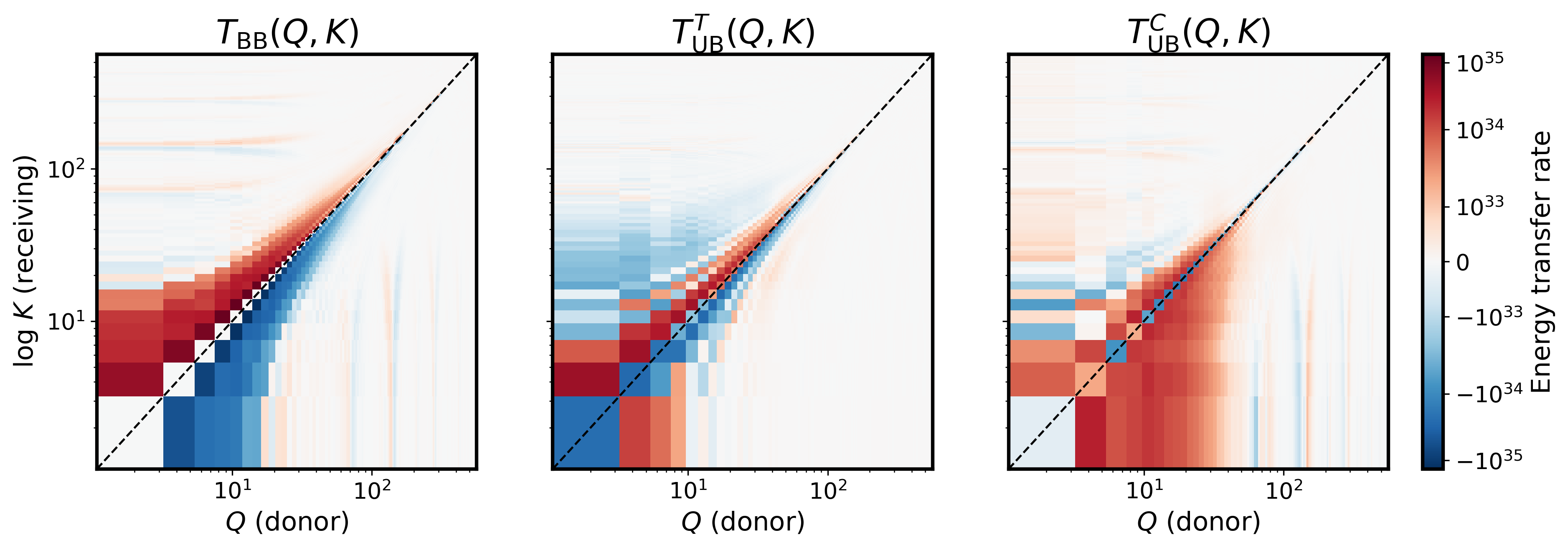}
    \includegraphics[width=0.8\linewidth]{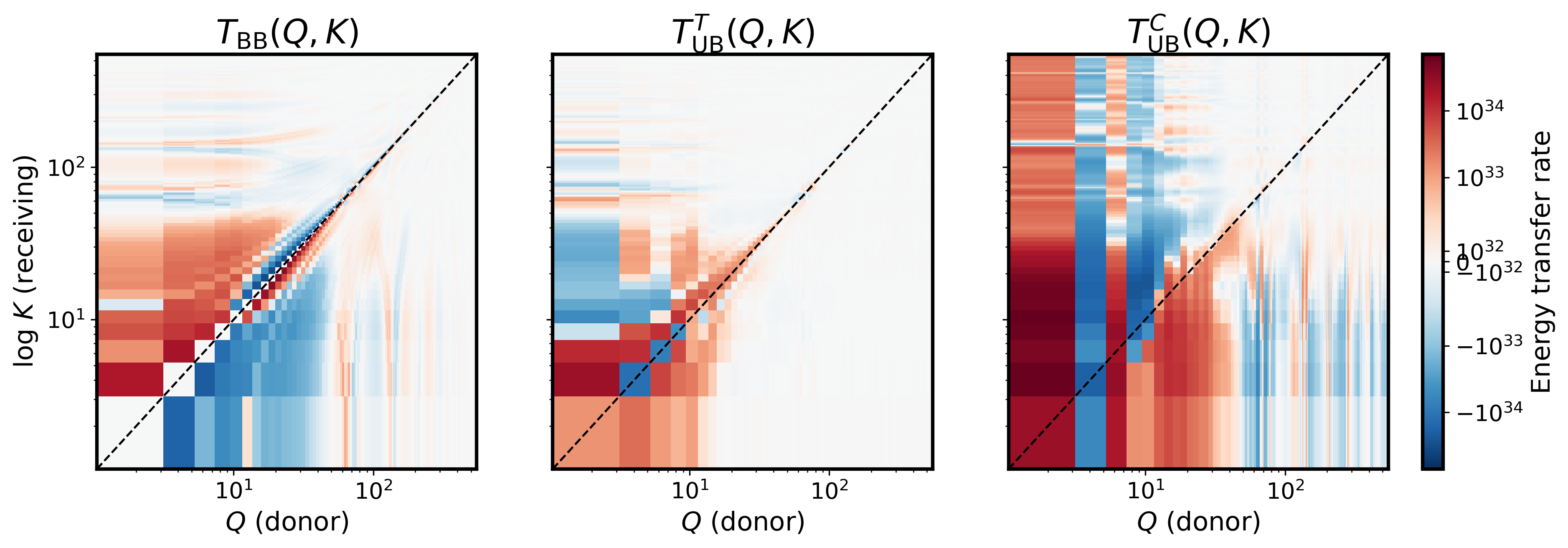}
    \includegraphics[width=0.8\linewidth]{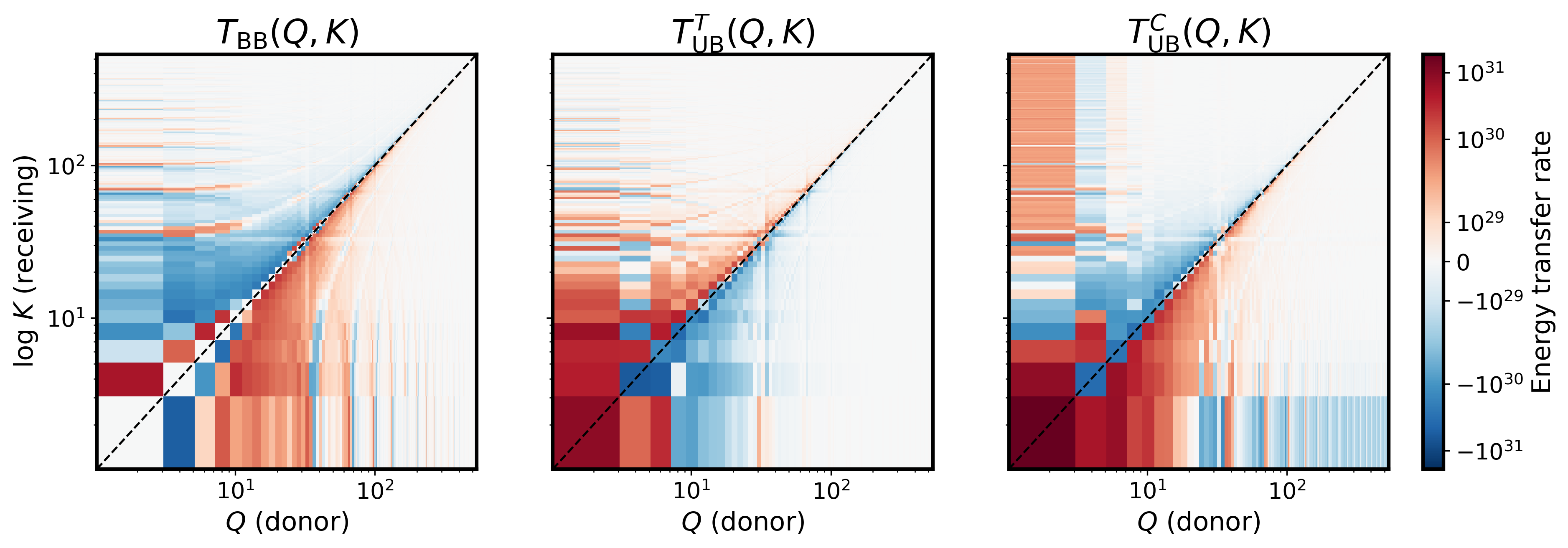}
    \caption{
    Spectral energy transfer rates, from magnetic to magnetic energy (left), from kinetic to magnetic energy mediated by tension (middle) and by compression (right) at $z=9.1,\, 8.9,\, 8.6$ (top to bottom). The x-axes are the donor and the y-axes the receiving wavelength, in inverse .
    }
     \label{fig:transfer_functions}
\end{figure*}

\begin{figure*}
    \centering
    \includegraphics[width=0.9\linewidth]{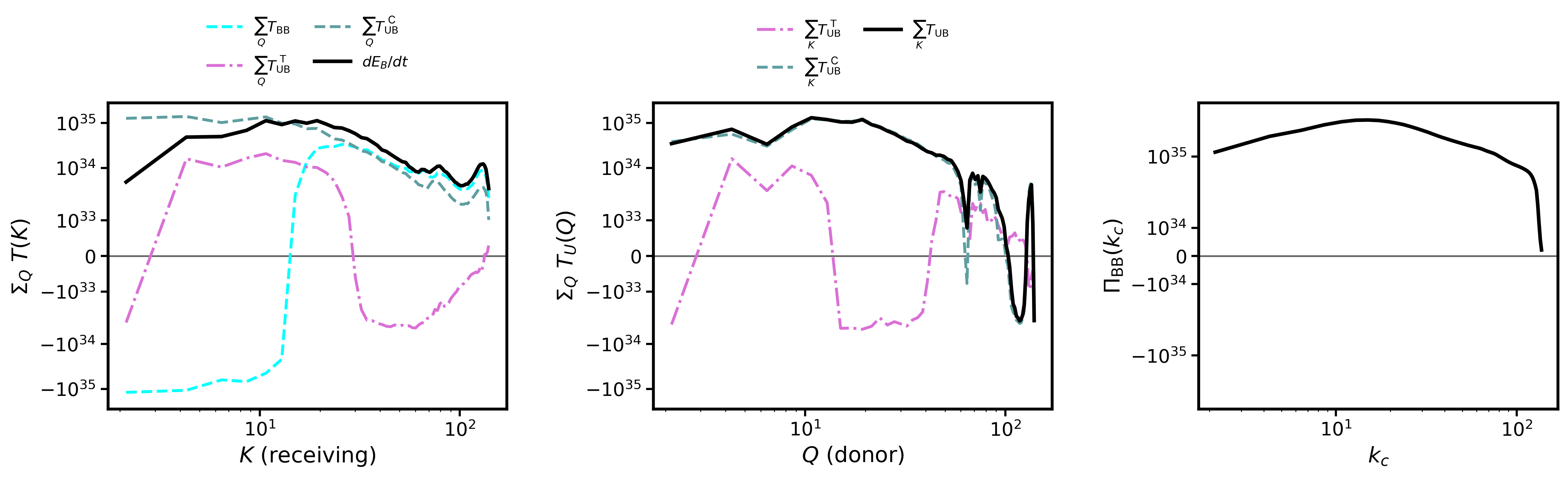}
    \includegraphics[width=0.9\linewidth]{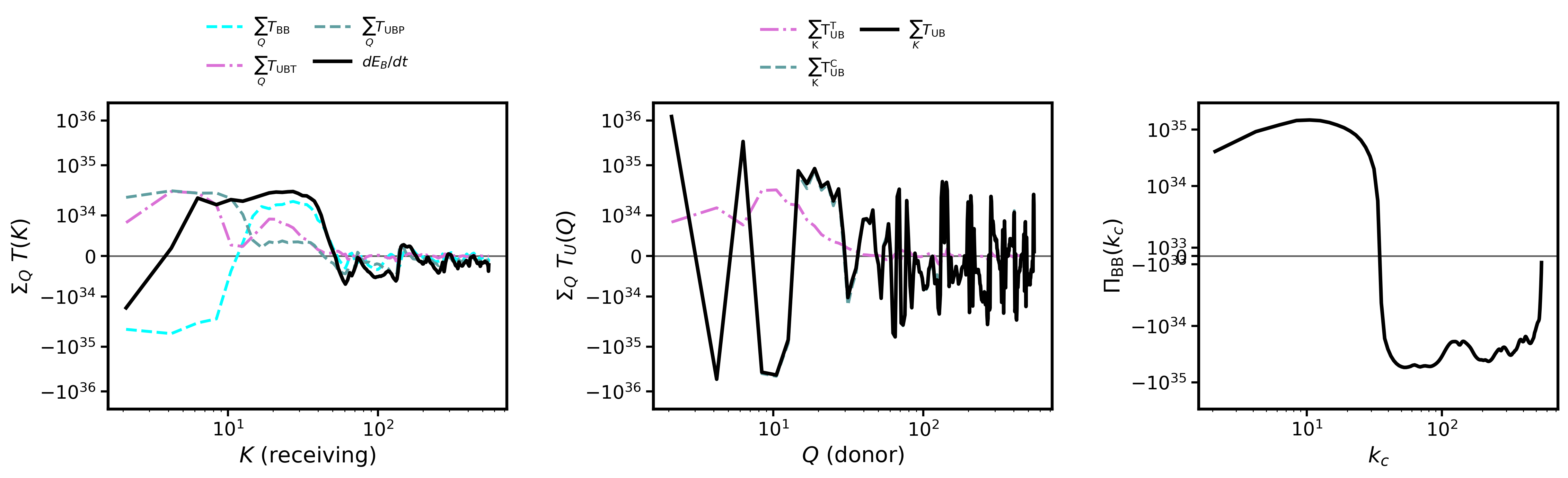}
    \includegraphics[width=0.9\linewidth]{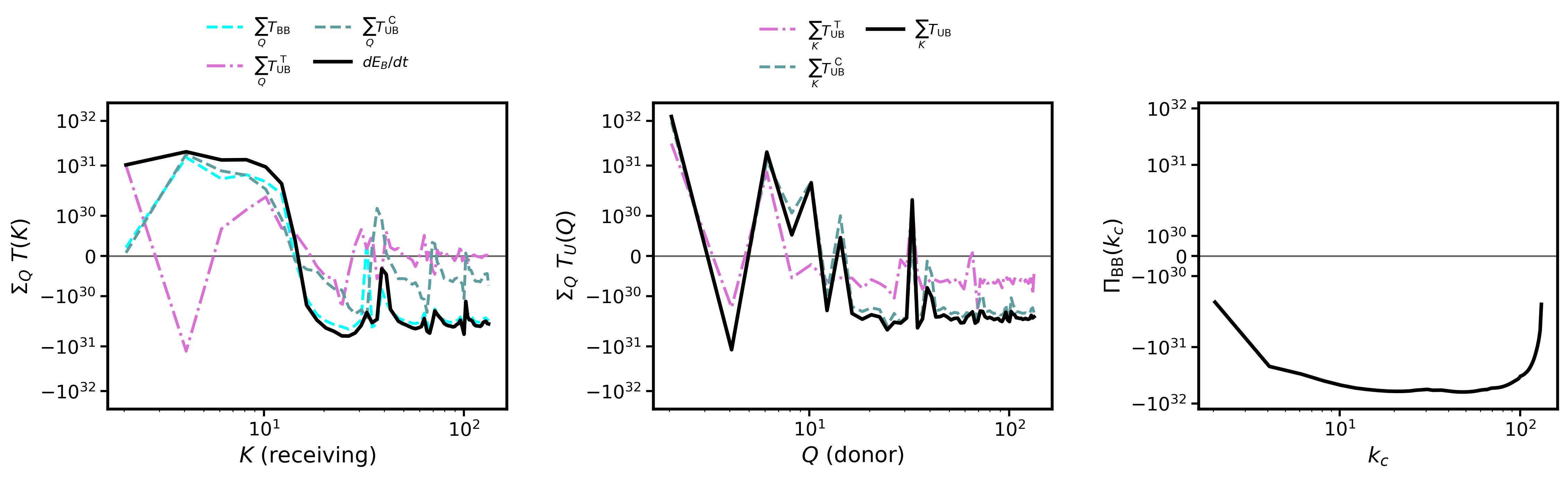}
    \caption{
    {\com{Spectral energy transfer rates per receiving shell (Eq. \ref{eq:transfer_K}, left) and per donor shell (Eq. \ref{eq:rec_shell}, middle), and total magnetic spectral flux across a cutoff wavenumber $\rm k_c$ (Eq. $\ref{eq:PiBB}$, right) at $z=9.1,\, 8.9,\, 8.6$ (top to bottom). $\rm T_{BB}$ denotes magnetic-to-magnetic energy conversion, $\rm T_{UB}^C$ kinetic-to-magnetic energy conversion through compression, and $\rm T_{UB}^T$ kinetic-to-magnetic energy conversion through magnetic tension. $\Pi_{BB}>0$ means a forward cascade and $\Pi_{BB}<0$ an inverse cascade.}}
    }
     \label{fig:1dtransfer_functions}
\end{figure*}

To gain some insight into the scale-dependent behavior of the magnetic field during its amplification in this very dynamic environment, we look at the time-dependent power spectra of the kinetic and magnetic energies in Fig. \ref{fig:power_spectra}.
{\com{The power spectra are computed after mapping the simulation quantities onto a uniform $256^3$ Cartesian grid enclosing the galaxy. 
Each Cartesian component of the resulting vector field is then Fourier transformed independently using a fast Fourier transform (FFT). For example, for the magnetic field we define the three-dimensional modal power as
\begin{equation}
    \rm P_B(\mathbf{k}) =
|\widehat{B}_x(\mathbf{k})|^2+
|\widehat{B}_y(\mathbf{k})|^2+
|\widehat{B}_z(\mathbf{k})|^2.
\end{equation}
Here and in the following we denote 
Fourier transforms with hatted symbols, here $\rm \widehat{B}_i$ for the Fourier transform of $i$th magnetic-field component. After applying the appropriate FFT normalization, the one-dimensional isotropic spectrum is obtained by binning the Fourier modes in spherical shells of constant radius $|\mathbf{k}|$ and taking the mean power of the modes contained in each shell.}}

{\com{Fig. \ref{fig:power_spectra}} shows that} both {\com{kinetic and magnetic}} power spectra are mostly flat on large scales, i.e. $k \lesssim 1/(100\,\rm pc)$.
The kinetic energy power spectra remain almost unaltered with time, with a steep slope equal to about $-11/3$, which is the Kolmogorov 3D power spectrum. The magnetic field power spectra instead flatten over time, with a slope close to $-5/3$, which corresponds to fully developed turbulence, at $k \gtrsim 1/(100\,\rm pc)$ after the strong amplification stage ($z=8.6$). This behavior clearly indicates that the magnetic field amplification is more pronounced on the smallest scales. The slopes in the inertial range of the spectra at ($z=8.6$) show how the environment has developed turbulence, and it is potentially favorable for a turbulent dynamo to act.

{\com{A more direct diagnostic of the physical origin and scale dependence of the magnetic field amplification is provided by the shell-to-shell spectral energy transfer formalism for compressible MHD \citep{Grete2017}. This formalism is very useful because it can describe the flow of energy within the magnetic reservoir and between the kinetic reservoir.

For any field $\mathbf{a}(\mathbf{x})$, its contribution from a Fourier shell $K$ is defined as
\begin{equation}
    \rm
    \mathbf{a}^{K}(\mathbf{x})
    =
    \sum_{\mathbf{k}\in\mathcal{S}_{K}}
    \widehat{\mathbf{a}}(\mathbf{k})
    \exp(i\mathbf{k}\cdot\mathbf{x}),
\end{equation}
where 
$\mathcal{S}_{K}$ is a spherical shell in Fourier space centered on the wavenumber $k_K$ and with a constant width $\rm \Delta k$. In the following, $\rm Q$ denotes the donor shell and $\rm K$ the receiving shell. Therefore, a positive transfer $\rm T(Q,K)$ corresponds to a net gain of energy by shell K from shell Q.

The conservative transfer of magnetic energy between magnetic shells is
\begin{equation}
\label{eq:TBB}
\rm     T_{\rm BB}(Q,K)
    =
    T_{\rm BB,adv}(Q,K)
    +
    T_{\rm BB,com}(Q,K),
\end{equation}
where
\begin{equation}
\label{eq:TBBa}
 \rm   T_{\rm BB,adv}(Q,K)
    =
    -\int
    \mathbf{B}^{K}\cdot
    \left[
    (\mathbf{u}\cdot\nabla)\mathbf{B}^{Q}
    \right]
    \,{\rm d}V
\end{equation}
is the advective part of the magnetic cascade, while
\begin{equation}
\label{eq:TBBc}
\rm    T_{\rm BB,com}(Q,K)
    =
    -\frac{1}{2}
    \int
    \left(
    \mathbf{B}^{K}\cdot\mathbf{B}^{Q}
    \right)
    \nabla\cdot\mathbf{u}
    \,{\rm d}V
\end{equation}
is the compressive correction required in the compressible case.
Their sum represents redistribution of energy within the magnetic reservoir and satisfies
\begin{equation}
   \rm T_{\rm BB}(Q,K)=-T_{\rm BB}(K,Q),
\end{equation}
so that the total magnetic energy is conserved.

The exchange between the kinetic and magnetic reservoirs is separated into contributions by magnetic tension and magnetic pressure. The tension-mediated transfer from kinetic shell Q to magnetic shell K is
\begin{equation}
\label{eq:TUBT}
 \rm   T_{\rm UB}^T(Q,K)
    =
    \int
    B_i^{K}
    \,
    \partial_j
    \left(
    v_{{\rm A},j}w_i^{Q}
    \right)
    \,{\rm d}V ,
\end{equation}
where $\rm v_A$ the Alfv\'{e}n speed, $\rm \mathbf{w}\equiv\sqrt{\rho}\mathbf{u}$ the density-weighted velocity, and repeated Cartesian indices imply summation. 
%
The corresponding magnetic pressure transfer is
\begin{equation}
\label{eq:TUBP}
    \rm T_{\rm UB}^C
    =
    -\int
    \left(
    \mathbf{B}^{K}\cdot\mathbf{B}
    \right)
    \nabla\cdot
    \left(
    \frac{\mathbf{w}^{Q}}{2\sqrt{\rho}}
    \right)
    \,{\rm d}V .
\end{equation}
This term quantifies how kinetic energy is transformed into magnetic energy through compressive work. We therefore refer to $\rm T_{UB}^C$ as the compressive amplification channel. For both $\rm T_{UB}^T$ and $\rm T_{UB}^C$, positive values indicate a net transfer from kinetic energy at the donor scale Q to magnetic energy at the receiving scale K, whereas negative values correspond to energy exchange in the opposite direction.

To quantify the net energy received by a magnetic scale K, we integrate over all donor shells.
\begin{equation}
\label{eq:transfer_K}
  \rm  T_X(K)=\sum_Q T_X(Q,K),
\end{equation}
where $\rm X\in\{{BB},{UB}^T,{UB}^C\}$. 
The resolved contribution to the magnetic-energy evolution at scale K is therefore
\begin{equation}
\label{eq:dEBdt}
   \rm \left.\frac{{\rm d}E_B(K)}{{\rm d}t}\right|_{\rm tr}
    =
    \sum_Q
    \left[
    T_{\rm BB}(Q,K)
    +
    T_{\rm UB}^T(Q,K)
    +
    T_{\rm UB}^C(Q,K)
    \right].
\end{equation}
Conversely, summing over receiving shells
%
\begin{equation}
    \rm T_X(Q)=\sum_K T_X(Q,K),
    \label{eq:rec_shell}
\end{equation}
identifies the donor scales that supply energy to the magnetic reservoir.

Finally, we characterize the direction of the magnetic cascade through the magnetic spectral flux across a cutoff wavenumber $\rm k_c$,
\begin{equation}
\label{eq:PiBB}
    \rm \Pi_{BB}(k_c)
    =
    \sum_{Q\leq k_c}
    \sum_{K>k_c}
    T_{BB}(Q,K).
\end{equation}
Here, $\rm \Pi_{BB}>0$ denotes an effective forward cascade, whereas $\rm \Pi_{BB}<0$ indicates an effective inverse cascade. 

In Fig.~\ref{fig:transfer_functions}, we show these transfer functions at three stages of the magnetic-field evolution, immediately before the amplification ($z=9.1$), close to its peak ($z=8.9$), and during the later amplified phase ($z=8.6$). The analysis is performed within the same 2~kpc field of view adopted for the kinetic and magnetic energy power spectra\footnote{During that time interval, the effective radius of the galaxy stays constant to about 0.8~kpc.}.
In Fig. \ref{fig:1dtransfer_functions} we show the integrated transfer functions and the net spectral flux (Eqs. \ref{eq:transfer_K}, \ref{eq:rec_shell}, \ref{eq:PiBB}) for the same snapshots and volume.

We see that, before the amplification, the transfer from kinetic to magnetic energy is dominated by the compressive channel ($\rm T_{UB}^C$) over a broad range of scales, whereas the tension-mediated contribution ($\rm T_{UB}^T$) is weaker and changes sign with scale. At the same time, $\rm \Pi_{BB}$ remains positive throughout the resolved range, showing that magnetic energy is redistributed predominantly toward higher wavenumbers. Thus, at the onset of amplification, compression supplies magnetic energy while a forward magnetic cascade transports energy to smaller scales.

Near the peak of the amplification, the energy transfer becomes more complex, as both figures illustrate. The magnetic pressure channel remains an important source of energy, but both channels (kinetic and magnetic) become increasingly scale-dependent. Simultaneously, $\rm \Pi_{BB}$ changes sign over a broad interval of wavenumbers, indicating the emergence of a net inverse magnetic transfer over part of the resolved range. We therefore interpret this stage as a transition from a predominantly forward magnetic cascade to a bidirectional redistribution of magnetic energy.

At the later stage, the magnetic spectral flux is negative over most of the resolved range. The corresponding shell-to-shell transfer maps show that magnetic energy is preferentially transferred from higher donor wavenumbers to lower receiving wavenumbers. This shows that, once a substantial magnetic field has been generated, its evolution is increasingly characterized by an inverse redistribution of magnetic energy toward larger spatial scales.}}

\section{Discussion and conclusions}

In this work, we have presented the first results from the MAGNA SERRA suite of simulations by studying Magnolia, a strongly magnetized galaxy with a mean magnetic field $\rm \langle B \rangle\simeq6\mu G$ and stellar mass $M_\star=1.68\times10^8~{\rm M}_\odot$ at redshift $z=8.6$. 

Magnolia's chaotic recent past, with clumps forming and merging over a few Myrs, and the fact that the magnetic energy strongly exceeds the non-thermal energy from feedback, imply that the magnetic field's amplification is not driven by a dynamo.
Further, the power spectra of the magnetic field are consistent with fully developed turbulence, with no indication of a Kazantzev-like spectrum.

A more detailed picture emerges from the magnetic energy spectral transfer functions. At the onset of the amplification, the dominant conversion from kinetic to magnetic energy is provided by compression, while the tension contribution is weaker and strongly scale-dependent. At the same time, the magnetic-to-magnetic energy transfer exhibits a positive spectral flux over the resolved range, indicating that part of the newly amplified magnetic energy is redistributed toward higher wavenumbers through a predominantly forward cascade. Near the peak of the amplification, the transfer pattern becomes bidirectional: compressive driving remains important, but the magnetic spectral flux changes sign over part of the spectrum, signalling the emergence of a net inverse transfer over an intermediate range of scales. By the end of the amplification, the magnetic flux is effectively negative over most of the resolved range, showing that the subsequent evolution is dominated by an upscale redistribution of magnetic energy from smaller to larger spatial scales.
Although a larger galaxy sample is needed to draw general conclusions, this first discovery already shows that dynamo action is not the only pathway to a strong galactic magnetic field.
{\com{Instead, we provide evidence that anisotropic compression can act as fast as a small-scale dynamo and include an inverse cascade of magnetic energy as a signature of the magnetic field's reorganization.}}

{\com{This work does not achieve the necessary resolution for capturing a small-scale dynamo. According to previous work \citep[e.g.,][]{gent2024}, a resolution of 2~pc or below is necessary in order to resolve the turbulent dynamo. The resolution achieved here is about 10~pc over the studied redshift range, which is far above that threshold. This resolution limitation supports an interpretation that does not involve dynamo action. Further, the limited data we have at a two times lower resolution (Fig. \ref{fig:energies_time}) show the same amplification trend as the reference run, which means that the mechanism at work is not close to the smallest scales achieved.
Given the available data, it is not possible to speculate on how the same system would evolve if the small-scale turbulence and associated dynamo were resolved, but it is an effort worth pursuing in follow-up work.
}}

Previous numerical work has shown significant magnetic field amplification in Milky Way progenitors and reported the magnetic field properties of galaxies at relatively low redshift (typically $z<2$). For example, \citet{martinalvarez2018, martinalvarez2022} studied the dynamo process in zoom-in cosmological galaxies,
and reported $\rm \mu G$ magnetic fields at $\rm z\simeq2$.
\citet{sparre2022} reported the effects of magnetic fields in cosmological mergers that reproduce Milky-Way-sized remnants at $\rm z=0$, with a focus on the gas flow around the galaxies. They found that the magnetic field in these galaxies reached equipartition with the turbulent kinetic energy. 
On larger scales, \citet{katz2021} studied the effects of cosmological magnetic fields on the 21~cm EoR signal, but did not focus on the galactic properties.

The present work reports for the first time a strongly magnetized galaxy ($\langle B \rangle\gtrsim1\, \rm\mu G$) already at redshift $z\simeq 9$ and almost at equipartition with thermal energy at $z\simeq8.6$. This finding is unexpected, since the galactic dynamo theory predicts slower magnetic field growth in galaxies (of the order of Gyr) driven by differential rotation \citep{Bran_EN23}.
This result is significant because it implies that galaxies can be magnetically dominated for almost their entire evolutionary history, with implications for star formation and cosmic ray propagation. 

While Magnolia is at a challenging redshift for observational detection, it is useful to consider the implications of its existence in light of recent high-redshift polarization detections.
Specifically, the first detections of polarized dust emission at $z=2.6$ and $z=5.6$ \citep{geach2023,chen2024} hint at dynamically important fields at those redshifts, which have put tension on theoretical models of galactic magnetism.
However, Magnolia appears with a dynamically important field almost half a Gyr earlier than the $z=5.6$ detection, comfortably accommodating the field growth implied by the observations. In addition, the large-scale order that we observe in Fig. \ref{fig:projection} indicates that a similarly organized magnetic field, given a strong dust content and enough magnification, could be detectable in dust polarization even with a physical beam of the size of hundreds of parsecs.
Interestingly, in the case of Magnolia, the signal would probably be beam-depolarized in the densest regions, but ordered and detectable in the intermediate-density regions around the star-forming clumps.

{\com{The current work has only explored one value and morphology for the initial field. Starting the simulations closer to the current observational lower limit of $B\gtrsim10^{-15}$~comoving G would probably delay the onset of such a strong field in the galaxy's ISM. Conversely, stronger fields might lead to an even earlier magnetization of the galaxy, or even a dynamical effect on its formation. The morphology of the initial field (uniform versus stochastic, for example, as in \citet{Konstantinou2024}) might also play a role. These interesting directions will be explored in follow-up simulations within the MAGNA SERRA suite.}}

While future analysis with proper dust radiative transfer is necessary to assess detectability, the large-scale, ordered magnetic field here means that systems similar to Magnolia are good candidates for future, magnified, high-$z$ dust-polarization detections.


\begin{acknowledgements}
 {\com{We thank the anonymous referee for a number of suggestions that greatly improved the quality of this manuscript and led to some new interpretations.}}
   We are also grateful to Andrea Ferrara for useful comments and suggestions during the early stages of this work.
   EN acknowledges funding from the Italian Ministry for Universities and Research (MUR) through the "Young Researchers" funding call (Project MSCA 000074). 
   We acknowledge projects AECT-2025-2-0051 (PI: EN) and AECT-2024-2-0003 (PI: FDS) at the MareNostrum machine of the Red Espa\~{n}ola de Supercoputaci\'{o}n (RES).
   We gratefully acknowledge the computational resources of the Center for High Performance Computing (CHPC) at SNS.
\end{acknowledgements}

\bibliographystyle{aa_url}
\bibliography{magnolia}


\end{document}